\documentclass[12pt]{iopart}

\usepackage{amssymb}
\newcommand{\ie}{\textit{i.e.}}
\newcommand{\eg}{\textit{e.g.}}

\def\pf{\mathop{\mathrm{Pf}}\nolimits}

\begin{document}
\title{Riemann-Liouville integrals of fractional order and extended KP hierarchy}
 
\author{Masaru Kamata\dag\  
\footnote[3]{e-mail address: nkamata@minato.kisarazu.ac.jp} 
and Atsushi Nakamula\ddag\
\footnote[4]{e-mail address: nakamula@sci.kitasato-u.ac.jp}}
\address{\dag\ Kisarazu National College of Technology,
    2-11-1 Kiyomidai-Higashi, Kisarazu, Chiba 292-0041, Japan}
\address{\ddag\ Department of Physics, School of Science, Kitasato University,
Sagamihara, Kanagawa 228-8555, Japan}

\begin{abstract}
An attempt is given to formulate the extensions of the KP hierarchy by introducing fractional order pseudo-differential operators.
In the case of the extension with the half-order pseudo-differential operators, a system analogous to the supersymmetric extensions of the KP hierarchy is obtained.
Unlike the supersymmetric extensions, no Grassmannian variable appears in the hierarchy considered here.
More general hierarchies constructed by the $1/N$-th order pseudo-differential operators, their integrability and the reduction procedure are also investigated.
In addition to finding out the new extensions of the KP hierarchy, brief introduction to the Riemann-Liouville integral is provided to yield a candidate for the fractional order pseudo-differential operators.
\end{abstract}

%\submitto{\JPA}

\maketitle

\section{Introduction}
Integrable hierarchies of nonlinear partial differential equations (PDE) have been vigorously studied from the perspective of physics as well as mathematics.
Among them the Kadomtsev-Petviashvili (KP) hierarchy and its variants appear in  many areas of theoretical physics.
In particular, the supersymmetric extensions of the KP hierarchy play important roles in non-perturbative superstring theories \cite{AIMZ}, and the connections are suggested between  the dispersionless limit of the KP hierarchy  and topological field theory \cite{AK,Dub}.

Concerning with the construction of the KP hierarchy in the Lax formalism, the non-commutative algebra of pseudo- or micro-differential operators play fundamental role \cite{DJKM}.
For the standard KP hierarchy, the associated pseudo-differential operator can be regarded as an ordinary integral operator, which enjoys the generalized Leibniz rule.
The aim of this paper is to inquire into the practicability of the extensions of the KP hierarchy by introducing fractional order pseudo-differential operators.
In this respect, we recall that the survey of the fractional order integration and differentiation is known as fractional calculus.

The fractional calculus, which usually stands for the differentiation and integration of arbitrary order so the terminology is somewhat misleading, has a long and rich history \cite{BW,SKM}.
The standard definition of the arbitrary order integration/differentiation is mostly given by the so called Riemann-Liouville integral these days.
Although the fractional calculus has been studied well in mathematics, it is not an ordinary mathematical tool in the theory of integrable systems at present.
Apart from integrable systems, there are many applications of fractional calculus in physics; for example, one of the present authors analyzed the supersymmetric field theories through half-order differential operators \cite{Kama}, other important works in the subject are performed on non-differential evolution equations, chaotic dynamical systems, material physics, and so on \cite{Hil}.

In the present paper we consider extensions of the KP hierarchy by introducing the fractional order integral/differential operators as pseudo-differential operators, which should be interpreted as the power roots of ordinary integration/differentiation; the situation is similar to the supersymmetric extensions of the KP hierarchy \cite{MR,Mula,Rab,UY}, where the square-root of integral/differential operators are brought in through superspace formulation.
In contrast, we extend the KP hierarchy by making use of fractional order integral/differential operators with respect to purely ``bosonic" variables, for which the relevant non-commutative algebra is the generalized Leibniz rule of fractional order.
We will see in the following that the extension of the KP hierarchy by half-order integral/differential operators leads to a hierarchy being similar to that of supersymmetric extension, as expected.

This paper is organized as follows.
In the next section we give a very brief review of the Lax formulation of the KP hierarchy and its supersymmetric extension for the purpose of determining notation.
In section 3 we make an attempt to generalize the KP hierarchy by fractional order integral/differential operators, and find the formulation works consistently.
In section 4 we introduce the Riemann-Liouville integrals as a candidate for the pseudo-differential operators of fractional order, which supply the generalized Leibniz rule being used in section 3.
The final section is devoted to concluding remarks.

\section{The Lax formulation of the KP hierarchy}
	In this section we give a sketch of the Lax formulation of the standard KP hierarchy, its $k$-reduction and supersymmetric extensions, to fix the notation throughout the present paper.

	\subsection{The standard KP hierarchy}
The KP hierarchy within the framework of Lax formulation is generated by the non-commutative algebra of the pseudo-differential operator $\partial^{-j}$ with respect to an independent variable $x$, which acts on a function through the generalized Leibniz rule,
\begin{equation}
\partial^{-j}\circ f=\sum_{k=0}^\infty\left(
{-j \atop k}
 \right)f^{(k)}\partial^{-j-k}.\label{Leibniz}
\end{equation}
Here we consider the case of integer $j$, the order of pseudo-derivative or integral, although the formula (\ref{Leibniz}) is valid for non-integer $j$. 
We define the Lax operator of the (one-component) KP hierarchy by,
	\begin{equation}
	\mathcal{L}_{KP}=\partial+\sum_{j=1}^\infty u_{j+1}\partial^{-j},\label{Laxop}
	\end{equation}
where $u_j$'s are dependent variables of space $x$ and time variables being introduced below.
The coefficient of $\partial^0$ can be set zero without loss of generality.
We assign the degree of the differential operator $\partial$ one, standing for $\deg[\partial]=1$, and assume that all the terms in the Lax operator (\ref{Laxop}) have equal degree, \ie, $\deg[u_j]=j$.
This assignment of the degree is naturally justified by the tau-function formalism.
Introducing infinite directions of ``time" $\mathbf{t}=(t_1,t_2,t_3,\dots)$ with $\mathrm{deg}[t_n]=-n$, we may consider the Lax equations,
\begin{equation}
\frac{\partial \mathcal{L}_{KP}}{\partial t_n}=[\mathcal{B}_n,\mathcal{L}_{KP}] \quad (n=1,2,3,\dots).\label{Laxeq}
\end{equation}
If we define the $n$-th ``Hamiltonian" $\mathcal{B}_n$ by the non-negative power part of $\partial$ in the $n$-th product of the Lax operator (\ref{Laxop}), denoting $\mathcal{B}_n:=(\mathcal{L}^n_{KP})_+=(\mathcal{L}^n_{KP})_{\geq0}$, we will obtain an infinite tower of nonlinear PDE's, the standard KP hierarchy.
Note that the lowest time variable $t_1$ should be identified with the space variable $x$ due to the first Lax equation.
The lowest PDE, the KP equation, is obtained by comparing the coefficients of $\partial^{-j}$'s in each side of (\ref{Laxeq}) for $t_2$ and $t_3$ developments of $u_2,\; u_3$ and $u_4$ and eliminating the $u_3$ and $u_4$,
\begin{equation}
\frac{3}{4}\frac{\partial^2 u}{\partial y^2}=
\left[\frac{\partial u}{\partial t}-\frac{1}{4}u'''-3uu'\right]'	,\label{KPeqn}
\end{equation}
where $u:=u_2$, $y:=t_2$ and $t:=t_3$, and the prime is the derivative with respect to $x(=t_1)$.

	\subsection{The $k$-reduction}
The KP hierarchy is an unconstrained system in the sense that the dependent variables $u_j$ are independent of each other in the Lax operator (\ref{Laxop}).
This independence is not necessary: we can impose  constraints between dependent variables without loss of consistency.
The most familiar is the $k$-reductions of the KP hierarchy for an integer $k\ge2$, for which the constraints are $\mathcal{L}^k_{KP}=\mathcal{B}_k$, \ie, all the coefficients of negative powers in $\partial$ of $\mathcal{L}^k_{KP}$ are zero, 
\begin{equation}
\left(\mathcal{L}^k_{KP}\right)_{-m}=0,\label{constraints}
\end{equation}
where $m=1,2,\dots$.
This is equivalent to the $t_{lk}$ independence of the system, for a natural number $l$.
For example, the $2$- and $3$-reduction lead to the KdV and the Boussinesq hierarchy, respectively.
For later purpose, we make a trivial remark that the reduction conditions (\ref{constraints}) are compatible with the Lax equation: the conditions  are invariant with respect to the time evolutions,
\begin{eqnarray}
\left(\frac{\partial \mathcal{L}^k_{KP}}{\partial t_n}\right)_{-m}
&=\left([\mathcal{B}_n,\mathcal{L}^k_{KP}]\right)_{-m}\nonumber\\
&=\left([\mathcal{B}_n,\mathcal{B}_k]\right)_{-m}\nonumber\\
&=0,\label{reduction}
\end{eqnarray}
because the Hamiltonians have only derivatives, \ie, non-negative power terms in $\partial$.

	\subsection{Supersymmetric extensions}
Supersymmetric extensions of the KP hierarchy (SKP) are vigorously studied by both mathematicians and physicists.
In particular, they appear in the context of superstring and/or quantum gravity theories \cite{AIMZ}.
The first supersymmetric extension was done by Manin and Radul \cite{MR}, referred to MRKP\footnote{Another formulation of SKP is given in \cite{Mula,Rab}.}, 
in which the differential operator in superspace, \ie, the superderivative, and its inverse,
\begin{equation}
D:=\frac{\partial}{\partial\theta}+\theta\frac{\partial}{\partial x},\label{sdiff}\quad D^{-1}=\theta+\frac{\partial}{\partial\theta}\left(\frac{\partial}{\partial x}\right)^{-1},
\end{equation}
play the parallel role of $\partial$ and $\partial^{-1}$ in the standard ``bosonic" KP.
Here, $\theta$ is a Grassmann odd variable, accordingly the square of $D$ turns out to be ordinary derivative,
\begin{equation}
D^2=\frac{\partial}{\partial x},\label{Dsquare}
\end{equation}
in other words $D$ can be regarded as a square root of $\partial$.
According to the superspace formalism, superfields $\Phi_j$ play the role of dependent variables in the MRKP, whose Lax operator of the MRKP is defined as,
\begin{equation}
\mathcal{L}_{MR}=D+\sum_{j=1}^\infty \Phi_jD^{1-j}.\label{MRLax}
\end{equation}
Besides the bosonic time variables $\mathbf{t}$, infinite fermionic time variables $(\tau_1,\tau_2,\dots)$ must be introduced.
Consequently we observe that both the bosonic and fermionic time flows of the superfields make up a system of super-differential equations.

To make a comparison with another extension of the KP hierarchy considered in the following section, we exhibit the lowest degree bosonic time flows of the MRKP, which is given by the Lax equation of even order,
\begin{equation}
\frac{\partial \mathcal{L}_{MR}}{\partial t_n}=[\mathcal{B}_{2n},\mathcal{L}_{MR}],\label{SKPLax}
\end{equation}
where the Hamiltonian is the standard one: $\mathcal{B}_{2n}:=(\mathcal{L}_{MR}^{2n})_+$.
In addition, there exist fermionic time flows given by odd order Lax equation certainly, we do not need them, however, in the present consideration, for the detail see \cite{MR,UY}.
One can show the lowest degree equation of (\ref{SKPLax}) is an extension of the KP equation (\ref{KPeqn}), which can be given in the component form \cite{BDPR}:
\numparts
\begin{eqnarray}
\frac{3}{4}\frac{\partial^2 u}{\partial y^2}=
\left[\frac{\partial u}{\partial t}-\frac{1}{4}u'''-3uu'+\frac{3}{2}v''v\right]',\label{MRKPu}\\
\frac{3}{4}\frac{\partial^2 v}{\partial y^2}=
\left[\frac{\partial v}{\partial t}-\frac{1}{4}v'''-\frac{3}{2}(uv)'
\right]',\label{MRKPv}
\end{eqnarray}\endnumparts
where the bosonic variable $u$ and the fermionic one $v$ are defined by $D\Phi_2=v+\theta u$, and $t:=t_3$ and $y:=t_2$.

Besides the MRKP, various types of supersymmetric extension of the KP hierarchy are considered \cite{ANP,NP}.
For example, a non-standard Lax equation by Brunelli and Das \cite{BD} leads to an extension of the KP equation of the following form,
\numparts
\begin{eqnarray}
	\frac{3}{4}\frac{\partial^2 u}{\partial y^2}=
\left[\frac{\partial u}{\partial t}-\frac{1}{4}u'''-3uu'-\frac{3}{2}v''v
-\frac{3}{2}v'\int^x\frac{\partial v}{\partial y}dx-\frac{3}{2}v\frac{\partial v}{\partial y}\right]',\label{BDKPu}\\
\frac{3}{4}\frac{\partial^2 v}{\partial y^2}=
\left[\frac{\partial v}{\partial t}-\frac{1}{4}v'''-\frac{3}{2}(uv)'-\frac{3}{2}u\int^x\frac{\partial v}{\partial y}dx+
\frac{3}{2}v'\int^x\frac{\partial u}{\partial y}dx
\right]',\label{BDKPv}
\end{eqnarray}\endnumparts
where, similarly to the MRKP, $u$ and $v$ are bosonic and fermionic variables, respectively.
In contrast to (\ref{MRKPu}) and (\ref{MRKPv}), there appear non-local terms in these coupled equations.

\section{Extensions of the KP hierarchy by fractional order integral operators}
	
This section provides the extensions of the standard KP hierarchy by fractional order integral operators, which is the main topic of the present paper.

	\subsection{Extended Lax operator}
Recall that the Leibniz rule (\ref{Leibniz}) is applicable when the order $j$ of ``integral" is an arbitrary real (or complex) number.
It will be interesting to consider the case when the Lax operator includes fractional order integrals, and then, to inquire whether the system gives a consistent hierarchy or not
\footnote{Here we restrict ourselves to rational $j$; if $j$ is irrational, the Lax equation could not give a closed system, see the following argument.}.
In the following consideration, we accept the axiom that the fractional order integral operators exist and also its exponential law $\partial^{-i}\partial^{-j}=\partial^{-(i+j)}$ holds for fractional $i$ and $j$, for a while.
We will see the Riemann-Liouville integral of fractional order enjoys these requirements in the next section.

		\subsubsection{Extension by the half-order integrals}
For the simplest case of an extension of the KP hierarchy, we consider the Lax operator including the half-order integrals in addition to the Lax operator (\ref{Laxop}).
We restrict ourselves to the case that the highest order term is $\partial$ as in the KP.
Accordingly, we define the most general half-order integral operator,
	\begin{equation}
	\mathcal{M}_{1/2}=v_3\;\partial^{-1/2}+v_5\;\partial^{-3/2}+v_7\;\partial^{-5/2}+\cdots,\label{Mhalf}
	\end{equation}
where $v_m$'s are the dependent variables of degree $m/2$.
We have set the ``differentiation" term $\partial^{1/2}$ to be absent: this resulted from the Lax equation defined below.
We remark that the Lax operator composed only of the half-order integrals (\ref{Mhalf}) itself does not produce any consistent hierarchy, because its products does not close in the half-order integral operators: we need integer order integral/differential operators to close the algebra.
With this definition, we consider the following Lax operator,
\begin{equation}
\mathcal{L}_{1/2}=\mathcal{L}_{KP}+\mathcal{M}_{1/2},\label{Lax2}
\end{equation}
and the standard Lax equation for the flows with respect to the time $\mathbf{t}=(t_1,t_2,\dots)$,
\begin{equation}
\frac{\partial \mathcal{L}_{1/2}}{\partial t_n}=[\mathcal{B}_n,\mathcal{L}_{1/2}].\label{Laxeq2}
\end{equation}
If we take the standard definition of the Hamiltonian, $\mathcal{B}_n:=(\mathcal{L}^n_{1/2})_+$, whose lower degree sequence is,
\numparts
\begin{eqnarray}
\mathcal{B}_1=\partial\\
\mathcal{B}_2=\partial^2+2v_3\;\partial^{1/2}+2u_2 \label{eKP2B2}\\
\mathcal{B}_3=\partial^3+3v_3\;\partial^{3/2}+3u_2\partial+3(v_5+v_3')\partial^{1/2}+3u_3+3u_2'+3v_3^2,\label{eKP2B3}
\end{eqnarray}
\endnumparts
then we find closed coupled nonlinear PDE's, an extended KP hierarchy by the half-order integrals, hereafter eKP$_{1/2}$.
Other ``non-standard" definitions of $\mathcal{B}_n$ such as $(\mathcal{L}^n_{1/2})_{\ge 1/2}$ cause inconsistency.
To show the consistency of the system, we derive the lowest degree coupled PDE from (\ref{Laxeq2}), the extended KP equation by the half-order integral, \ie, the eKP$_{1/2}$ equation.
Just like the original KP equation (\ref{KPeqn}), we need the first two non-trivial equations of (\ref{Laxeq2}).
Each coefficient of the negative powers in $\partial$ of
\begin{equation}
\frac{\partial \mathcal{L}_{1/2}}{\partial t_2}=[\mathcal{B}_2,\mathcal{L}_{1/2}],
\end{equation}
is,
\numparts
\begin{eqnarray}
&\partial^{-1/2}:& \frac{\partial v_3}{\partial y}=2v'_5+v''_3\\
&\partial^{-1}:& \frac{\partial u_2}{\partial y}=2u_3'+u_2''+2v_3v_3'\\
&\partial^{-3/2}:& \frac{\partial v_5}{\partial y}=2v_7'+v_5''+2(v_3u_2)'\\
&\partial^{-2}:& \frac{\partial u_3}{\partial y}=2u_4'+u_3''+2u_2u_2'+3v_5v_3'+v_3v_5'-v_3v_3'',
\end{eqnarray}\endnumparts
whereas of
\begin{equation}
\frac{\partial \mathcal{L}_{1/2}}{\partial t_3}=[\mathcal{B}_3,\mathcal{L}_{1/2}],
\end{equation}
is,
\numparts
\begin{eqnarray}
&\partial^{-1/2}:& \frac{\partial v_3}{\partial t}=3v'_7+3v_5''+v'''_3+6(v_3u_2)'\\
&\partial^{-1}:& \frac{\partial u_2}{\partial t}=3u_4'+3u_3''+u_2'''+6u_2u_2'+6(v_3v_5)'\nonumber\\
&&\qquad+\frac{3}{2}(v_3'^2+v_3v_3'').
\end{eqnarray}\endnumparts
Eliminating the dependent variables $u_3,\; u_4,\; v_5$ and $v_7$, we find the coupled PDE with non-local term,
\numparts
\begin{eqnarray}
\frac{3}{4}\frac{\partial^2 u}{\partial y^2}=
\left[\frac{\partial u}{\partial t}-\frac{1}{4}u'''-3uu'
+\frac{3}{8}(v^2)''-\frac{3}{4}v'\int^x\frac{\partial v}{\partial y}dx-\frac{3}{4}v\frac{\partial v}{\partial y}
\right]',\label{KP2u}\\
\frac{3}{4}\frac{\partial^2 v}{\partial y^2}=
\left[\frac{\partial v}{\partial t}-\frac{1}{4}v'''-3(uv)'
\right]',\label{KP2v}
\end{eqnarray}\endnumparts
where $u:=u_2$ and $v:=v_3$.
As expected, (\ref{KP2u}) reduces to the KP equation (\ref{KPeqn}) when $v$ is absent.
We observe the resemblance between (\ref{KP2u}), (\ref{KP2v}) and the MRKP equations (\ref{MRKPu}), (\ref{MRKPv}) or the non-standard SKP equations (\ref{BDKPu}), (\ref{BDKPv}), however they are not exactly identical.
This resemblance obviously comes from the fact that the derivative in superspace can be read as a square root of the derivative, which fact is formally equivalent to the feature of the half order derivative $\partial^{1/2}$.
In contrast to the supersymmetric extensions, the extension considered in this section works without using Grassmann numbers.

		\subsubsection{Extension by the $1/N$-th order integrals}
Having observed the extension by the half-order integrals is successful, we now consider more generic extensions by the $1/N$-th order integrals ($N=3,4,\dots$), eKP$_{1/N}$ hierarchies.
In these cases, we need to introduce integral operators $\partial^{-1/N},\;\partial^{-2/N},\;\dots,\partial^{-(N-1)/N}$ simultaneously to give a consistent Lax equation, since we have to close the commutator algebra in the Lax equations under the axiom $\partial^{-i}\partial^{-j}=\partial^{-i-j}$.
For $N=p$, a prime number, there appears a new system coupled to the KP hierarchy.
For example, we give an outline of the $N=3$ case, in which the Lax operator should be made up of, 
\begin{equation}
\mathcal{L}_{1/3}=\mathcal{L}_{KP}+\mathcal{M}_{1/3}+\mathcal{M}_{2/3},
\end{equation}
 where,
\begin{eqnarray}
	\mathcal{M}_{1/3}=w_4\;\partial^{-1/3}+w_7\;\partial^{-4/3}+w_{10}\;\partial^{-7/3}+\cdots,\label{M1/3}\\
	\mathcal{M}_{2/3}=w_5\;\partial^{-2/3}+w_8\;\partial^{-5/3}+w_{11}\;\partial^{-8/3}+\cdots\label{M2/3},
\end{eqnarray}
and $\deg [w_m]=m/3$.
We observe that the standard Lax equation and the definition of the Hamiltonian  similar to the former case give a consistent hierarchy of coupled PDE's.
One can see the lowest coupled PDE arises from the first two non-trivial Lax equations.
Each coefficient of $\partial$ in 
\begin{equation}
\frac{\partial \mathcal{L}_{1/3}}{\partial t_2}=[\mathcal{B}_2,\mathcal{L}_{1/3}],
\end{equation}
is,
\numparts
\begin{eqnarray}
&\partial^{-1/3}:& \frac{\partial w_4}{\partial y}=2w'_7+w''_4\\
&\partial^{-2/3}:& \frac{\partial w_5}{\partial y}=2w'_8+w''_5+(w_4^2)'\\
&\partial^{-1}:& \frac{\partial u_2}{\partial y}=2u_3'+u_2''+2(w_4w_5)'\\
&\partial^{-4/3}:& \frac{\partial w_7}{\partial y}=2w'_{10}+w''_7+(w_5^2)'+2(w_4u_2)'\\
&\partial^{-5/3}:& \frac{\partial w_8}{\partial y}=2w_{11}'+w_8''+\frac{8}{3}w_7w_4'+\frac{4}{3}w_4w_7'+2w_5u_2'-\frac{2}{3}w_4w_4''\\
&\partial^{-2}:& \frac{\partial u_3}{\partial y}=2u_4'+u_3''+2u_2u_2'+\frac{10}{3}w_8w_4'+\frac{4}{3}w_4w_8'+\frac{8}{3}w_7w_5'\nonumber\\
&&\qquad+\frac{2}{3}w_5w_7'-\frac{4}{3}w_5w_4''-\frac{2}{3}w_4w_5'',
\end{eqnarray}\endnumparts
whereas of 
\begin{equation}
\frac{\partial \mathcal{L}_{1/3}}{\partial t_3}=[\mathcal{B}_3,\mathcal{L}_{1/3}],
\end{equation}
is,
\numparts
\begin{eqnarray}
&\partial^{-1/3}:& \frac{\partial w_4}{\partial t}=3w'_{10}+3w_7''+w'''_4+6(w_4u_2)'+3(w_5^2)'\\
&\partial^{-2/3}:& \frac{\partial w_5}{\partial t}=3w'_{11}+3w_8''+w'''_5+6(w_5u_2)'+6(w_7w_4)'\nonumber\\
&&\qquad+2w_4'^2+2w_4w_4''\\
&\partial^{-1}:& \frac{\partial u_2}{\partial t}=3u_4'+3u_3''+u_2'''+6u_2u_2'+6(w_8w_4)'+6(w_7w_5)'\nonumber\\
&&\qquad+w_5w_4''+3w_5'w_4'+3w_4^2w_4'.
\end{eqnarray}\endnumparts
These are nine equations for the nine dependent variables so that we can combine them into the coupled PDE of $u_2,\; w_4$ and $w_5$.

For $N$ being a composite number, we observe that the new system is coupled to the system coming from the prime factors of $N$.
For example, the eKP$_{1/4}$ system is a new system coupling to the eKP$_{1/2}$ system given above.

Finally we should remark that the introduction of pseudo-derivative of irrational order does not make a finite closed system: we need uncountable number of additional $\mathcal{M}$'s like (\ref{M1/3}) and (\ref{M2/3}).

\subsection{The conservation laws}
Since the eKP$_{1/N}$ hierarchy is constructed within the framework of Lax formalism, we expect that the system is  integrable \textit{a priori}.
In fact, we observe that there are infinite conservation laws, which can be derived by standard procedure \cite{Segal} for the Lax operator under consideration, say $\mathcal{L}_*$, then we find,
\begin{equation}
\frac{\partial}{\partial t_m}\mathrm{Res}(\mathcal{L}_*^{\;n})=P_{m,n}',\label{CC}
\end{equation}
where the residue is defined as $(\mathcal{L}_*^{\;n})_{-1}$ and $P_{m,n}$ is a differential polynomial of $u_j$ and $v_j$.
One can see that the presence of the fractional order integral operators do not modify the formula (\ref{CC}).
Hence we expect the existence of many special solutions to the eKP$_{1/N}$ hierarchies, just like the solitons in the original KP.

In addition to these conserved charges with integer degree, we have another set which has non-integer degree.
For concreteness, we consider the eKP$_{1/2}$, in which there exist conserved charges with degree $k+1/2\; (k=0,1,2,\dots)$: we can find the charges come from,
\begin{equation}
\mathrm{Res}(\mathcal{L}_{1/2}^{k+\frac{1}{2}}),
\end{equation}
where the square root of the Lax operator is constructed by usual procedure,
\begin{equation}
\mathcal{L}_{1/2}^{\ \frac{1}{2}}=\partial^{\frac{1}{2}}+\frac{1}{2}v_3\partial^{-1}+\frac{1}{2}u_2\partial^{-\frac{3}{2}}+\frac{1}{2}(v_5-\frac{1}{4}v_3')\partial^{-2}+\cdots.
\end{equation}
Although we can construct the charges with half-integer degree, there does not exist a consistent time flow generated by the Hamiltonian $\mathcal{B}_{k+\frac{1}{2}}:=\mathcal{L}_{1/2}^{k+\frac{1}{2}}$.

In general, for the eKP$_{1/N}$ $(N\geq3)$ we will find the existence of additional sequences of conserved charges.

\subsection{The $k$-reduction of the eKP$_{1/N}$}
In this subsection we make a comment on the $k$-reduction of the eKP$_{1/N}$, the truncation of the $t_{lk}$ flow.
Unlike the standard KP hierarchy, the reduction condition  $\mathcal{L}^k_{1/N}=\mathcal{B}_k$ does not work in the eKP$_{1/N}$ hierarchies due to the property of fractional integrals.
For, the compatibility (\ref{reduction}) between the reduction condition and the Lax equation  does not hold when the fractional order ``derivative" operators are present in the Hamiltonian $\mathcal{B}_k$.
This comes from the fact that the Leibniz rule for non-integer order is not a finite sum even if the order $j$ is positive, hence the right hand side of the corresponding equation to (\ref{reduction}) induces negative terms in $\partial$, \ie,
\begin{equation}
\left([\mathcal{B}_k,\mathcal{B}_{k'}]\right)_{-\frac{m}{N}}\neq0,\quad (m=1,2,3,\dots).
\end{equation}
For example, the commutator of the Hamiltonians $\mathcal{B}_2$ and $\mathcal{B}_3$ for the eKP$_{1/2}$ gives,
\begin{equation}
\left([\mathcal{B}_3,\mathcal{B}_2]\right)_{-\frac{1}{2}}=\frac{3}{2}(v_3u'_2)'-(v_3^3)'\neq0,
\end{equation}
as well as the coefficient of $\partial^{-m/2}$ $(m=2,3,\dots)$.
Although it is not clear at present whether one can impose consistent reduction conditions on the Lax operator of the eKP$_{1/N}$, we may consider the truncated system by hand.

Here we present a remarkable fact that there exists an algebraic solution to the $2$-reduction of the eKP$_{1/2}$ equation (\ref{KP2u}) and (\ref{KP2v}), which should be referred eKdV$_{1/2}$ equation,
\numparts
\begin{eqnarray}
\frac{\partial u}{\partial t}=
\frac{1}{4}u'''+3uu'-\frac{3}{8}(v^2)''\label{KdV2u},\\
\frac{\partial v}{\partial t}=\frac{1}{4}v'''+3(uv)'.\label{KdV2v}
\end{eqnarray}\endnumparts
Notice the strong resemblance between the supersymmetric KdV equation \cite{Math} and (\ref{KdV2u}), (\ref{KdV2v}).
The solution is an extension of the rational solution to the KdV equation:
\numparts
\begin{eqnarray}
u(x,t)=-\frac{5}{16}\frac{1}{(x+ct)^2}+\frac{c}{3}, \\
v(x,t)=\pm\frac{\sqrt{165}}{24}\frac{1}{(x+ct)^{3/2}},
\end{eqnarray}\endnumparts
where $c$ is a constant with $\deg[c]=2$, which is required by keeping the correct degree of the dependent variables.

\section{The Riemann-Liouville integrals of fractional order}

So far we constructed the extensions of the KP hierarchy by introducing the pseudo-differential operator of fractional order, the fractional integral.
However, we have treated such operators as only generators of a non-commutative algebra.
In the construction of the standard KP hierarchy, we use the Leibniz rule of negative order derivatives (\ref{Leibniz}), which can be realized by the iterative use of integration by parts, \eg, for $j=1$,
\begin{eqnarray}
\partial^{-1}(fg)=\int^x fg\;dx&=fG^{(1)}-\int^x f'G^{(1)}dx\nonumber\\
&=fG^{(1)}-f'G^{(2)}+\int^x f''G^{(2)}dx\nonumber\\
&=fG^{(1)}-f'G^{(2)}+f''G^{(3)}-\cdots \nonumber\\
&=\sum_{k=0}^\infty(-1)^k f^{(k)}G^{(k+1)},\label{Leibniz-1}
\end{eqnarray}
where $G^{(k)}$ is $k$-times indefinite integral of $g$.
The cases of higher order $j$ can be shown by multiplicative operation of (\ref{Leibniz-1}).
Hence we accept the statement that the negative order derivative operator is equivalent to indefinite integral operator.

For the case of fractional $j$, how can we realize the formula (\ref{Leibniz})?
To find out the appropriate fractional integration/derivation on a function assumed to be existing in the last section, we recall the Riemann-Liouville integral of order $\alpha\in\mathbb{C}$ for $\Re e\;\alpha>0$, 
\begin{equation}
I^\alpha f(x):=
\frac{1}{\Gamma(\alpha)}\int_a^x(x-z)^{\alpha-1}f(z)dz,\label{RL}
\end{equation}
where $f(x)$ is assumed to be locally integrable and rapidly decreasing on the lower boundary $a$.
We realize that when $\Re e\;\alpha<0,\; \alpha\notin -\mathbb{N}$,  (\ref{RL}) should be read as,
\begin{equation}
I^\alpha f(x):=I^{\alpha+n}f^{(n)}(x)=\frac{d^n}{dx^n}I^{\alpha+n} f(x),\label{RLnega}
\end{equation}
where $n$ is the first integer of $n+\Re e\,\alpha>0$.
The assumption for $f(x)$ guarantees the last equality in (\ref{RLnega}), which means that the Riemann-Liouville integral commutes with the ordinary derivative.
Hence we understand that the definition (\ref{RL}) can be extended by analytic continuation to the whole $\alpha$, for the detail see Appendix.
One can also show, when $\alpha$ tends to an integer, the Riemann-Liouville integral turns out to be the ordinary integration/differentiation, as it should be.
In addition, we will observe in Appendix that the Riemann-Liouville integral complies with the exponential or additive index law, $I^\alpha I^\beta=I^{\alpha+\beta}$, which is assumed in the derivation of the eKP$_{1/N}$ hierarchy.

We exhibit the action of the Riemann-Liouville integral on some functions,
\begin{equation}
I^\alpha e^{\lambda x}=\frac{1}{\lambda^\alpha}e^{\lambda x}, \label{Iexp}
\end{equation}
where $\lambda >0$ and,
\begin{equation}
I^\alpha x^\mu=\frac{\Gamma(\mu+1)}{\Gamma(\alpha+\mu+1)}x^{\mu+\alpha}, \label{Ixmu}
\end{equation}
where the operand, $x^\mu$, is defined zero if $x<0$.

In Appendix we observe that the definitions (\ref{RL}) and (\ref{RLnega}) lead to the Leibniz rule (\ref{Leibniz}) for fractional order, 
\begin{equation}
I^\alpha \left(fg\right)=\sum_{k=0}^\infty\left({-\alpha \atop k}\right)f^{(k)}(I^{k+\alpha}g),\label{Leibnizfrac}
\end{equation}
where
\begin{equation}
\left({-\alpha \atop k}\right)=(-1)^k\frac{\Gamma(\alpha+k)}{\Gamma(\alpha)\;k!}.
\end{equation}
Note that (\ref{Leibnizfrac}) turns out to be (\ref{Leibniz-1}) as $\alpha$ tends to $1$, obviously.
In this respect, we should remark the fact that the Riemann-Liouville integrals do not define uniquely the Leibniz rule (\ref{Leibnizfrac}): further generalization to the rule is also possible \cite{BW}.
However it is sufficient for the present purpose that the Riemann-Liouville integrals can fulfil (\ref{Leibnizfrac}).
With the features observed above, we accept that the Riemann-Liouville integrals yield the pseudo-differential operators of fractional order,
\begin{equation}
\partial^{-\alpha}=I^\alpha.
\end{equation}
Thus we find that the pseudo-differential operator of fractional order in the Lax operator of the eKP$_{1/N}$ hierarchy is not only an element of a non-commutative algebra but a concrete operator on a certain class of functions.

Although we do not have needed the explicit operation of the fractional integrals in the derivation of the extended KP hierarchies considered in the previous section, we expect that the direct application of the Riemann-Liouville integral (\ref{RL}) is required in further consideration of the eKP$_{1/N}$ system.
For, we will need the Riemann-Liouville integrals when we consider the analytic solutions to the eKP$_{1/N}$ hierarchies through, \eg, the inverse scattering method, B\"acklund transformation, Painlev\'e analysis and so on \cite{Cont}.

\section{Concluding remarks}

In this paper we have considered the extensions of the KP hierarchy by introducing fractional order integral/differential operators, the eKP$_{1/N}$.
In particular, if we introduce the half-order integral operator, we find the resulted eKP$_{1/2}$ hierarchy is analogous to the SKP hierarchies.
Other extensions, the eKP$_{1/N}$ hierarchies, are also considered and the fact that there exist infinite conserved currents is observed.
We have also found an algebraic solution to the eKdV$_{1/2}$ equation, the reduced eKP$_{1/2}$ equation.

To obtain the deep understanding to the eKP$_{1/N}$ hierarchies, it is profitable to make an analysis in the Sato formulation, and also the tau-function formulation, just as in the case of the standard KP hierarchy and the SKP hierarchies \cite{IMM,MM,LiuMan}. 
Apart from the whole hierarchies, it will be interesting to investigate simply the integrability of the coupled equations like (\ref{KP2u}) and (\ref{KP2v}) through the Painlev\'e analysis \cite{KNT}.
Another approach is also attainable: the Hirota bilinear method is applicable to find special solutions like solitons.
As mentioned in the last section, it will be necessary to consider the Riemann-Liouville integrals explicitly for these analysis.

Incidentally, apart from integrable systems there are many works on fractional order evolution equations for relaxation, diffusion, stochastic process and so on \cite{Hil}.
Although we have obtained the nonlinear PDE's with normal derivative through the application of fractional calculus, it is intriguing to formulate a systematic procedure for creating the PDE's with anomalous derivative, \ie, fractional order derivative.
A possibility will be given by the application of another ``Leibniz rule" in the formulation given in this paper, in fact, the Riemann-Liouville integrals enjoy miscellaneous types of ``Leibniz rule" as mentioned in the last section.

\appendix
\setcounter{section}{1}
\section*{Appendix}

%The derivation of the exponential law.
This appendix is devoted to present some of the important properties of the Riemann-Liouville integrals (\ref{RL}).

First of all, we observe the exponential law $I^\alpha I^\beta=I^{\alpha+\beta}$.
For $\Re e\, \alpha,\; \Re e\,\beta>0$,
\begin{eqnarray}
I^\alpha I^\beta f(x)&=\frac{1}{\Gamma(\alpha)}\int_a^x dz(x-z)^{\alpha-1}\frac{1}{\Gamma(\beta)}\int_a^z dw (z-w)^{\beta-1}f(w)&\nonumber\\
&=\frac{1}{\Gamma(\alpha)\Gamma(\beta)}\int_a^x dwf(w) \int_z^x  dz(x-z)^{\alpha-1} (z-w)^{\beta-1}\nonumber\\
&=\frac{1}{\Gamma(\alpha+\beta)}\int_a^x dw(x-w)^{\alpha+\beta-1}f(w)\nonumber\\
&=I^{\alpha+\beta}f(x),\label{explaw}
\end{eqnarray}
where we changed the integration order and used the fact that the $z$-integral in the second line was expressed by the Beta function.
This formula can be extended to the case when the one or both of $\Re e\,\alpha$ and $\Re e\,\beta$ is negative, by (\ref{RLnega}) and the commutativity of $I^\alpha$ and ordinary derivative.

Next we derive the Leibniz rule for fractional order integral/differential operator (\ref{Leibnizfrac}).
For $\Re e\,\alpha>0$, if we expand one of the operands, say, $f$ in Taylor series, we find,
\begin{eqnarray}
I^\alpha(f(x)g(x))&=\frac{1}{\Gamma(\alpha)}\int_a^x(x-z)^{\alpha-1}f(z)g(z)dz&\nonumber\\
&=\frac{1}{\Gamma(\alpha)}\int_a^x(x-z)^{\alpha-1}\sum_{k=0}^\infty \frac{(-1)^{k}}{k!}f^{(k)}(x)(x-z)^k g(z)dz\nonumber\\
&=\frac{1}{\Gamma(\alpha)}\sum_{k=0}^\infty \frac{(-1)^{k}}{k!}f^{(k)}(x)\int_a^x(x-z)^{k+\alpha-1} g(z)dz\nonumber\\
&=\frac{1}{\Gamma(\alpha)}\sum_{k=0}^\infty \frac{(-1)^{k}}{k!}f^{(k)}(x)\Gamma(k+\alpha)I^{k+\alpha}g(x)\nonumber\\
&=\sum_{k=0}^\infty \left({-\alpha \atop k}\right)f^{(k)}(x)I^{k+\alpha}g(x),
\end{eqnarray}
which can be extended to $\Re e\,\alpha<0$ by ordinary differentiation (\ref{RLnega}).

Finally we comment on the Riemann-Liouville integral of order $\Re e\,\alpha<0$, \ie, a differentiation of arbitrary order.
In the definition (\ref{RLnega}) the integral is well-defined, however if we simply put $\Re e\,\alpha<0$ in the definition of $I^\alpha$ (\ref{RL}), then the definition turns out to be divergent integral whatever the function $f$ is, due to the singularity of the kernel $(x-z)^{\alpha-1}$ at the upper bound.
To give a well-defined meaning for the divergent integral, we can take the finite part of it, the $\pf$ (\textit{partie finie}) prescription.
To see this  we consider only the case $-1<\Re e\,\alpha<0$ for simplicity, the case of $\Re e\,\alpha<-1$ can be treated similarly.
We now define the ``regularized integral" with a positive cutoff parameter $\epsilon$ as,
\begin{eqnarray}
I^\alpha_\epsilon f&(x):=\frac{1}{\Gamma(\alpha)}\int_a^{x-\epsilon}(x-z)^{\alpha-1}f(z)dz\nonumber\\
&=\frac{1}{\Gamma(\alpha)}\int_\epsilon^{x-a}s^{\alpha-1}f(x-s)ds\nonumber\\
&=\frac{1}{\Gamma(\alpha+1)}\left\{s^\alpha f(x-s)\Big|_\epsilon^{x-a}+\int_\epsilon^{x-a}s^{\alpha}f'(x-s)ds\right\}\nonumber\\
&=\frac{1}{\Gamma(\alpha+1)}\left\{-\epsilon^\alpha f(x-\epsilon)+\int_\epsilon^{x-a}s^{\alpha}f'(x-s)ds\right\}\nonumber\\
&=\frac{1}{\Gamma(\alpha+1)}\left\{-\epsilon^\alpha \sum_{k=0}^\infty \frac{(-1)^k}{k!}f^{(k)}(x)\epsilon^k+\int_\epsilon^{x-a}s^{\alpha}f'(x-s)ds\right\},\label{regularization}
\end{eqnarray}
due to the condition $f(a)=0$ and $\Gamma(\alpha+1)=\alpha\Gamma(\alpha)$.
In the last line of (\ref{regularization}), when we take $\epsilon\to0$, the terms of $\epsilon^{\alpha+k}$ in the infinite sum are zero if $k\ge1$, and also the integral term remains finite. 
Thus we can perform the $\pf$ prescription by the following definition,
\begin{eqnarray}
I^\alpha f(x)&=\frac{1}{\Gamma(\alpha)}\pf\int_a^x(x-z)^{\alpha-1}f(z)dz\nonumber\\
&:=\lim_{\epsilon\to0}\left\{I^\alpha_\epsilon f(x)-\frac{(-1)}{\Gamma(\alpha+1)}\epsilon^\alpha f(x)\right\}\nonumber\\
&=\frac{1}{\Gamma(\alpha+1)}\int_0^{x-a}s^\alpha f'(x-s)ds\nonumber\\
&=I^{\alpha+1}f'(x).
\end{eqnarray}
We therefore conclude that the definition (\ref{RLnega}) is well-defined and the Riemann-Liouville integral can be continued analytically to negative $\Re e\,\alpha$.

\end{document}